\documentclass[letterpaper,superscriptaddress,aps,twocolumn]{revtex4-2}
\usepackage{amsmath,amssymb}
\usepackage{bm}
\usepackage{graphicx,color,natbib}
\usepackage{placeins}
\usepackage[dvipsnames]{xcolor}
\usepackage{siunitx}
\usepackage{oplotsymbl}

\newcommand{\beq}{\begin{equation}}
\newcommand{\eeq}{\end{equation}}
\newcommand{\beqa}{\begin{eqnarray}}
\newcommand{\eeqa}{\end{eqnarray}}

\newcommand{\bfr}{{\bf r}}

\newcommand{\bfv}{{\bf v}}
\newcommand{\bfu}{{\bf u}}

\newcommand{\bmS}{{\bm S}}

\newcommand{\bphi}{{\bm \phi}}
\newcommand{\bbeta}{{\bm \bbeta}}

\newcommand{\rvec}{\hat{\bfr}}
\newcommand{\phivec}{\hat{\bphi}}

\newcommand{\mum}{\mu\mathrm{m}}

\newcommand{\EQ}{E_{\mathrm{Q}}}

\newcommand{\PO}{\Pi_{\varphi}}

\newcommand{\Nmax}{N_{\mathrm{max}}}

\definecolor{strawberry}{rgb}{1.0,0.0,0.5}
\newcommand{\paddyspeaks}[1]{{\color{black} #1}}

\newcommand{\pecl}{\operatorname{\mathrm{P\kern-.08em e}}}

\begin{document}

\title{Complex flow profiles in microscopic active crystals}

\author{Abraham Mauleon-Amieva}
\affiliation{H.H. Wills Physics Laboratory, Tyndall Avenue, Bristol, BS8 1TL, UK}
\affiliation{School of Chemistry, University of Bristol, Cantock's Close, Bristol, BS8 1TS, UK}
\affiliation{Centre for Nanoscience and Quantum Information, Tyndall Avenue, Bristol, BS8 1FD, UK}
\affiliation{Bristol Centre for Functional Nanomaterials, Tyndall Avenue, Bristol, BS8 1FD, UK}

\author{Tanniemola B. Liverpool}
\affiliation{School of Mathematics, University of Bristol, Bristol, BS8 1TW, UK}

\author{Ian Williams}
\affiliation{School of Mathematics and Physics, University of Surrey, Guildford GU2 7XH, UK}

\author{Anton Souslov}
\affiliation{Department of Physics, University of Bath, Claverton Down, Bath, BA2 7AY}

\author{C. Patrick Royall}
\affiliation{H.H. Wills Physics Laboratory, Tyndall Avenue, Bristol, BS8 1TL, UK}
\affiliation{School of Chemistry, University of Bristol, Cantock's Close, Bristol, BS8 1TS, UK}
\affiliation{Centre for Nanoscience and Quantum Information, Tyndall Avenue, Bristol, BS8 1FD, UK}

\begin{abstract}
Active solids emerge from self-actuating components interacting with each other 
to form crystalline patterns. In equilibrium, commensurability underpins our understanding of nanoscale friction and particle-level dynamics of crystals. However, these concepts have yet to be imported into the realm of active matter. Here, we develop an experimental platform and a theoretical description for microscopic clusters composed of active particles confined and self-assembled into small crystals. In our experiments, these crystallites form upon circular confinement of active rollers, with a magic number of 61 rollers per well. Competition between solidity and self--propulsion leads to self--shearing and complex flow--inversion behaviour, along with self--sliding states and activity--induced melting. We discover active stick--slip dynamics, which periodically switch between a commensurate static state and an incommensurate self-sliding state characterised by a train of localised defects. We describe the steady--state behaviour using a discretised model of active hydrodynamics. We then quantify the intermittent stick--slip dynamics using a self-propelled extension of the Frenkel-Kontorova (FK) model, a fundamental workhorse of slipping and flow in crystals. Our findings in a colloidal model system point to a wealth of phenomena in incommensurate active solids as design principles for both assembly and robotics down to the nanoscale.
\end{abstract}

\maketitle

\section{Introduction}
Active self--assembly goes beyond the minimisation of free energy to design exotic structures and dynamics. Perhaps the simplest self--assembled structures are crystallites. When created from active components, crystalline clusters exhibit complex assembly based on either external stimuli~\cite{palacci2013} or internal self--propulsion~\cite{tan2022}.
Once assembled, crystallite behaviour remains profoundly affected by the underlying activity of individual building blocks. For example, crystal structure interplays with active dynamics, leading to characteristic patterns of self--kneading~\cite{bililign2022} and synchronization of collective modes~\cite{baconnier2022}. However, the physical principles behind active crystallites have only recently begun to be explored.

Self-organisation in active matter takes many forms, from flock formation~\cite{marchetti2013,bricard2013,bechinger2016}, to clustering through motility-induced phase separation~\cite{fily2012,buttinoni2013}. One way to control self-assembly is confinement, which has been developed as a workhorse from the molecular to the colloidal scale by perturbing systems at their boundaries~\cite{albasimionesco2006,lowen2009,williams2013}. Additional pathways of controlling these systems can be accessed by taking them out of mechanical equilibrium by supplying energy from the boundaries, \emph{e.g.}, by shearing~\cite{williams2016,williams2022,ortizambriz2018}. However, an exciting
way to take a system out of equilibrium is by using active components, which supply energy throughout the bulk rather than exclusively at boundaries or surfaces.

\begin{figure*}
\centering
\includegraphics[width=0.99\textwidth]{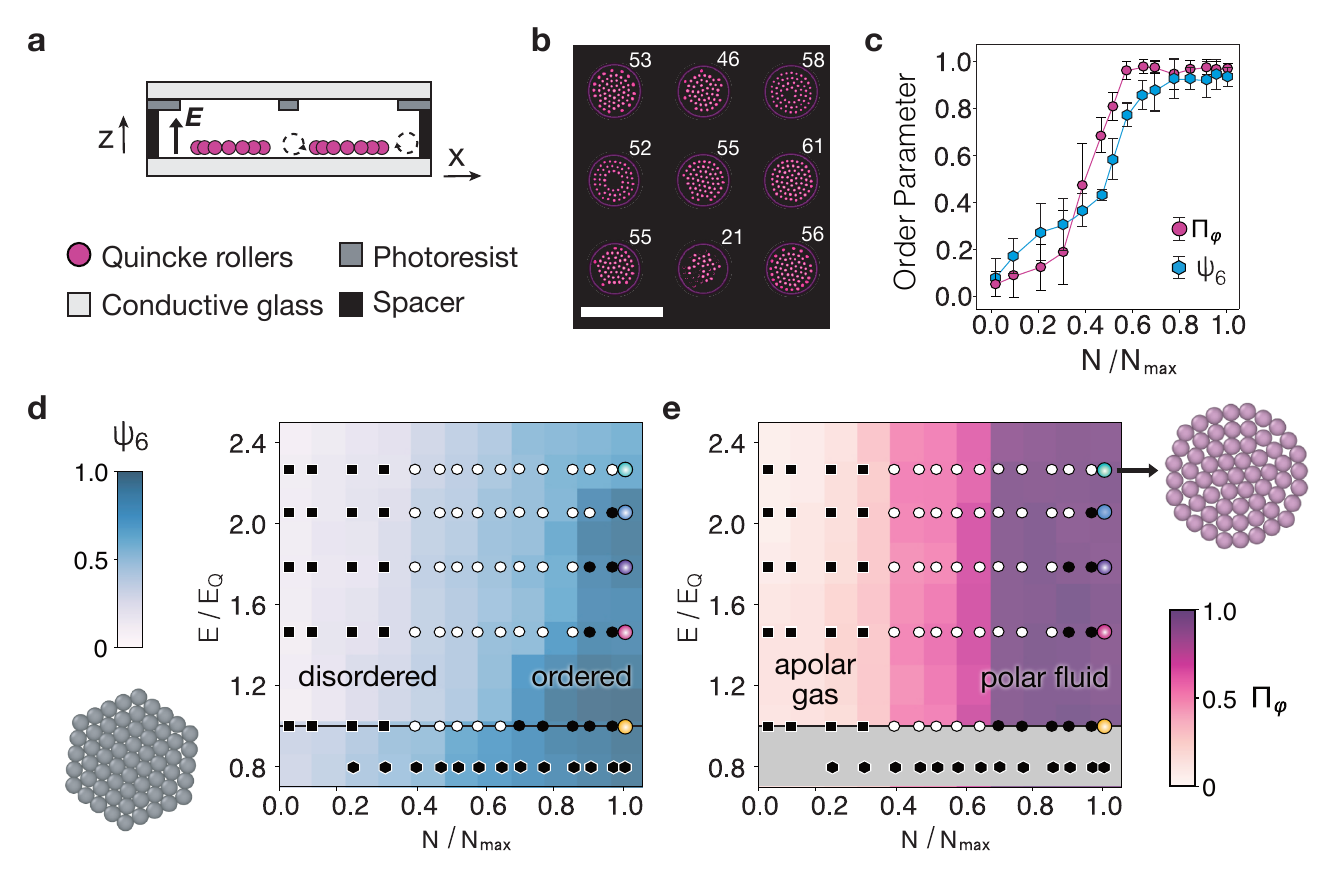}
\caption{\textbf{Confinement induced self--assembly of active Quincke rotors.}
\textbf{a.} The confinement of Quincke rollers into circular regions is achieved by the application of a localised electric field $E$.
\textbf{b.} Different populations $N$, indicated by the labels, are found across the confining regions. Scale bar represents $50\, \mu m$.
\textbf{c.} Polarisation $\PO$ and bond orientational order, $\psi_{6}$, parameters measured for different populations as they become active, \emph{i.e.} at $E \approx \EQ$.
\textbf{d--e.} Phase diagrams indicating (\textbf{d}) the hexagonal $\psi_{6}$ and (\textbf{e}) polar $\PO$ order for passive and active rollers under strong confinement. Above $\EQ$ (solid horizontal line), clusters become active and give rise to apolar (disordered) gas and polar (ordered) fluid phases. Colourbars indicate the magnitude of order parameters. Symbols are: (hexagon) passive aggregates, active (squares) apolar ($\PO \le 0.5$) and (circles) polar states ($\PO > 0.5$). Open white circles indicate polar fluids ($\psi_6 \le 0.5$), whereas solid black circles indicate active crystallites ($\psi_6 > 0.5$). Colours in data points for $N_\mathrm{max}$ correspond to data in Fig.~\ref{figN61}.}
\label{figSetUp}
\end{figure*}

Active solids exhibit pattern formation, and collective dynamics not seen in active fluids. Living systems present many examples of active solids, from collections of organisms~\cite{tan2022}, gastrulation~\cite{henkes2020}, to confluent tissues~\cite{park2014}.
These biological systems are complex, and considerable insight can be gained from examples of these exotic phenomena in a minimally simple setting. To this end, tunable colloidal model systems provide a means to realize such behavior in a wide variety of situations~\cite{bechinger2016}. However, increasing the packing fraction such that self-propelled colloidal systems solidify has proven challenging due to limitations in colloidal stability. Among the relatively few examples that have been studied are active colloidal glasses~\cite{klongvessa2019,geyer2019}. Addressing the challenges of working with active colloids at high density has allowed us to perform experiments to explore many of the open questions about active colloidal crystals.

Here, we use an experimental colloidal system to explore how the competition between crystalline order and self--propulsion leads to self--shearing of active crystallites. We characterise the activity-induced rotation, melting, and complex flow behaviour due to the dynamic frustration between self--propulsion and collective rotation. We observe that increasing self--propulsion leads to a sequence of rigid--body rotation, faster angular velocity at the edge and, finally, an inversion to faster angular velocity at the centre. We quantify this behaviour by discretising a version of the Toner--Tu equations. Using a minimal model of self--sliding, we connect our active clusters to the Frenkel--Kontorova (FK) model and nanotribological phenomena~\cite{persson,bohlein2012}.
In particular, slipping is associated with a breakdown in local crystalline order. We expect these active analogues of sliding friction to be generically present in active colloids that form crystallites, due to the frustration between particle motility and periodic order.

\section{Experimental set-up}
We address the challenge of stablising active colloids at high density with the time--honoured colloid science method of steric stabilisation. In particular, we use a suspension of active colloidal rollers powered by the Quincke electro--rotation mechanism~\cite{quincke1896}. Spontaneous rotation of colloids emerges with the application of a DC electric field, $E$, with amplitude above a critical value, $\EQ$. When adjacent to a surface, these particles roll in a random direction, as rotation couples with translation~\cite{bricard2013}, forming a quasi-two-dimensional ensemble of active particles. Our particular system uses colloids of diameter $\sigma = 2.92\,\mu$m dispersed in a solvent with low dielectric constant 
containing an ionic surfactant (See Methods for more details). The particles are laterally confined by applying the electric field only in a circular region of radius $R_{\mathrm{c}} \approx 5 \sigma$ (Fig.~\ref{figSetUp}\textbf{a}). At low amplitudes of the field, \emph{i.e.}, $E < \EQ$, where $\EQ \approx 0.8\, \mathrm{V}\,\mu \mathrm{m}^{-1}$, a population of $N$ colloids moves towards the confining regions and assembles into clusters. Figure~\ref{figSetUp}\textbf{b} shows nine confining regions, each with a different population $N$. The maximum population observed is $\Nmax = 61$. Full experimental details may be found in the Methods and the nature of the dynamics is shown in Supplementary Movie 1.

\begin{figure*}
\centering
\includegraphics[width=0.99\textwidth]{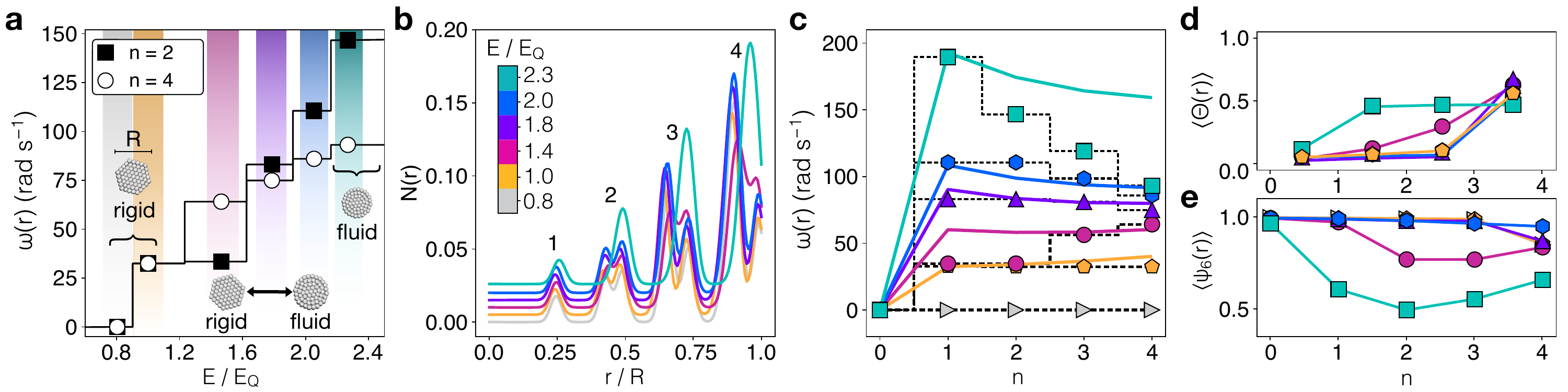}
\caption{\textbf{The magic number rotor with} $N = \Nmax = 61$.
\textbf{a.} Time-averaged angular velocity as a function of $E$ in layers $n=2$ (black squares) and $n=4$ (white circles), with $n$ counting outwards from the central particle (labelled $n = 0$). At intermediate field strength $E \approx 1.5 \EQ$ the outer layer rotates faster than the inner layer, but this situation is inverted at high $E > 1.6 \EQ$, where the inner layers rotate faster than the outer layer. Colours correspond to the state points shown in Figs.~\ref{figSetUp}\textbf{d--e}.
\textbf{b.} Radial density profiles characterising structures obtained at different fields $E$, indicated in the legend. Numbers over the peaks indicate the layer number $n$. Curves are offset vertically for clarity.
\textbf{c.} Angular velocity profiles as function of the layer number $n$, where $n=0$ is the central particle. Symbols are experimental data, solid lines are fits obtained using Eq.~5 in the SI, and dotted lines are step guides.
\textbf{d.} Distortion order parameter, $\langle \Theta \rangle$, as a function of layer $n$.
\textbf{e.} Hexagonal order parameter, $\langle \psi_{6} \rangle$, as function of layer $n$. For all panels, colours correspond to the electric field values indicated by the colourbar in~\textbf{b}.}
\label{figN61}
\end{figure*}


\section{Phase Behaviour in the Population--Activity plane}
The structural and dynamic behaviour in the population--activity ($N$--$E$) plane is summarised in Fig.~\ref{figSetUp}\textbf{b--e}. For $E < \EQ$, passive, hexagonally ordered crystallites are formed due to the electro--hydrodynamic interactions induced by the application of $E$, as detailed in Ref.~\cite{mauleon2020}. To elucidate the local hexagonal order, we use the particle--averaged bond orientational order parameter
$\psi_{6} = \langle \frac{1}{N}\sum_{j}^{N} |\psi_{6}^{j}| \rangle_{t}$, where
$\psi_{6}^j \equiv \frac{1}{z_{j}}\sum_{k=1}^{z_{j}}\exp(i6\theta_{k}^{j})$ quantifies the local order. The quantity $z_{j}$ is the co--ordination number of particle $j$ from a Voronoi tessellation, and $\theta_{k}^{j}$ is the angle between the bond from $i$ to $k$ and a reference axis. The parameter $\psi_{6}^j$ runs between perfect ordering ($\psi_{6} = 1$) and complete disorder ($\psi_{6} =0$). The collective dynamics are characterised by the time-averaged polar order parameter, $\PO = \langle | \frac{1}{N} \sum_{i}^{N} (\mathbf{\hat{u}}_{i} \cdot \hat{\mathbf{e}}_{\varphi})| \rangle_{t}$, which quantifies the degree of alignment between rollers. Here $\mathbf{\hat{u}}_{i}$ is the local orientation of roller $i$, and $\hat{\mathbf{e}}_{\varphi}$ is a unit vector along the azimuthal direction. $\PO = 1$ indicates perfect azimuthal alignment, and $\PO = 0$ represents random particle orientations.

Figure~\ref{figSetUp}\textbf{c} shows the dynamical order parameter $\PO$ compared to $\psi_{6}$ as a function of $N$ at the onset of instability, $E \approx \EQ$. A transition from theisotropic gas to a polar state is seen from the polarisation order parameter $\PO$, even for populations with decreasing structural order, \emph{i.e.}, $N \approx 0.5 \Nmax$. This is accompanied by an analogous increase in hexagonal order $\psi_6$. These two concurrent signatures indicate the emergence of an active crystallite at high densities.

In Fig.~\ref{figSetUp}\textbf{d}, we consider the hexagonal order parameter $\psi_6$ in the $(N,E)$ plane. For $E < \EQ$, \emph{i.e.}, passive systems, we see moderate values of $\psi_6$ even at low populations, $N/\Nmax<0.5$, corresponding to the formation of inactive clusters formed due to electrohydrodynamic attractions. In these passive systems, $\psi_6$ increases with population, and clusters become increasingly crystalline for $N$ approaching $\Nmax$. Switching on activity by increasing the field strength to $E > \EQ$ leads to a drop in $\psi_6$ as hexagonal ordering competes with activity~\cite{mauleon2020}. For the highest field strengths of $E/\EQ=2.3$, we observe the formation of a layered fluid structure in which particles are organised in concentric layers \cite{williams2016} (see right side of Fig. \ref{figSetUp}\textbf{e}). 

At the onset of Quincke rotation ($E = \EQ$), the role of density becomes apparent. With population $N \lesssim 0.5 \Nmax$, activity causes hexagonal ordering $\psi_{6}$ to decrease, but for larger populations, $\psi_{6}$ remains at values similar to those of passive crystallites.

Figure~\ref{figSetUp}\textbf{e} depicts the experimental phase diagram obtained from the polar order parameter, $\PO$, in the same $(N,E)$ plane. The passive and active phases are separated by the instability at field $E = \EQ$. Below $\EQ$, the polar order parameter is close to zero, 
however, when the system is active, we find a transition between an apolar gas (low $\PO$) at low population to a polar fluid (high $\PO$) as $N$ increases. The boundary between these states is insensitive to field strength. Comparing Figs.~\ref{figSetUp}\textbf{d} and \textbf{e}, we find two ordered regimes: a polar fluid at medium density ($N/\Nmax \ge 0.4$) and an active crystalline phase at even higher densities in which both parameters $\PO$ and $\psi_6$ approach $1$. At the highest densities, we observe that sufficiently strong activity is still able to melt the crystalline order, a phenomenon we now turn to.


\section{Steady--State Slipping: Activity vs the Boundary}

To investigate the interplay between crystalline order and dynamics, we focus on a specific population, $N = \Nmax = 61$, and  the activity--dependent rotational dynamics of both rigid--hexagonal and layered--fluid structures. In its quiescent state for $E \lesssim \EQ$, this population forms a nearly perfect hexagonal shape, see Fig.~\ref{figSetUp}. At high field strengths, \emph{i.e.}, $E \approx 2.3 \EQ$, the structure is fluid--like with four concentric layers around a central stationary particle. The active state exhibits field--dependent structures with different dynamics, see Fig.~\ref{figN61}\textbf{a}. For both hexagonal and layered--fluid cases, the different layers are well resolved by the radial density profiles $N(r)$ in Fig.~\ref{figN61}\textbf{b}. One signature of hexagonal structure is a split peak which continuously merges into a single peak as the system becomes more fluidized. 


Upon increasing activity $E$, we observe a range of dynamical scenarios, including rigid--body rotation and periodic rearrangements, as illustrated by the angular velocity, $\omega$, profiles in Fig.~\ref{figN61}\textbf{c}. In particular, in order of ascending field strength:
(\emph{i}), for $E < \EQ$, no rotation is measured (grey triangles);
(\emph{ii}), at the onset of activity, we observe rigid--body rotation (yellow pentagons);
(\emph{iii}), the cluster rotates fastest at the outside  (magenta circles);
(\emph{iv}), we find rigid--body rotation (purple triangles);
(\emph{v}), the cluster rotates fastest in the interior (blue hexagons and teal squares).
Therefore, we find an \emph{inversion} in the flow profile as a function of field strength. This complex dynamical behaviour in Fig.~\ref{figN61}\textbf{c} contrasts with our time--averaged measure of structure, Fig.~\ref{figN61}\textbf{b}, which shows little change until the very highest field strength. The bond--orientational parameter $\psi_6$ is a more sensitive measure of structure (Fig.~\ref{figN61}\textbf{e}), and shows additional detail such as a drop in hexagonal order when the outer layers rotate faster (magenta data). 
However, the lack of sensitivity of these time--averaged structural measures motivate a time--resolved approach, which we pursue below.

\begin{figure}
\centering
\includegraphics[width=0.475\textwidth]{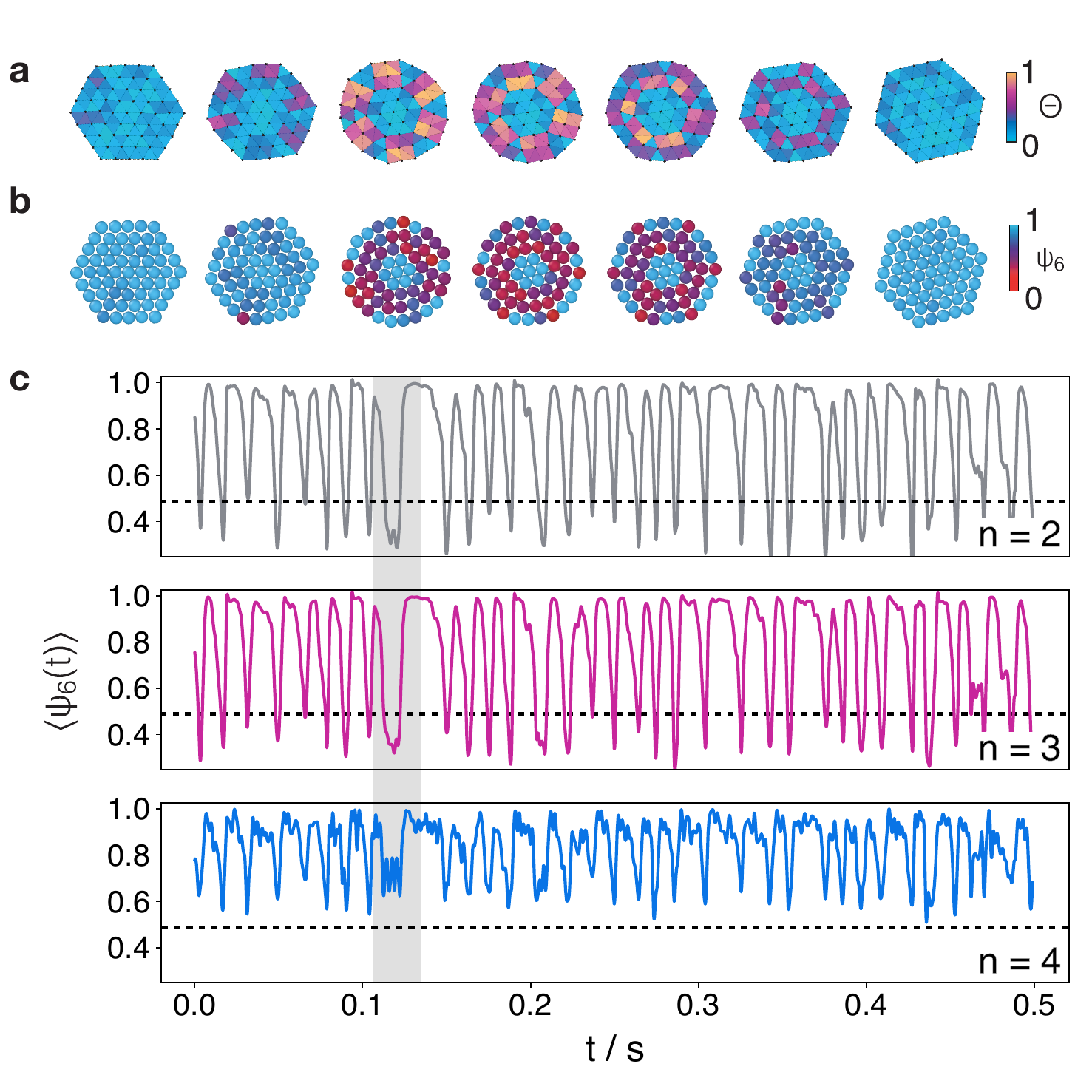}
\caption{\textbf{Slipping layers and order evolution.} \textbf{a--b.} Time sequence for a single layer--slipping event, showing the change of the distortion parameter $\Theta$ (\textbf{a}) and the hexagonal order parameter $\psi_{6}$ (\textbf{b}). \textbf{c.} Evolution of $\psi_{6}$ in layers $n = 2$, $3$, and $4$. Incommensurate and commensurate configurations are indicated by the loss and recovery of $\psi_{6}$. Horizontal dashed lines indicate $\psi_{6} = 0.5$. Shaded region corresponds to the sequence in \textbf{a--b}.}
\label{figFK}
\end{figure}

Before exploring the time--resolved behaviour, we now consider a minimal model to describe the steady-state behaviour~\cite{toner1995}. This model has two key ingredients:
(\emph{i}) a constant preferred self-propulsion velocity for an individual particle layer in isolation (captured by preferred speed $u_0$ and friction $\xi$ that speeds up or slows down a particle towards the preferred speed) and (\emph{ii}) a sliding friction between adjacent layers (captured by viscosity $\tilde\nu$), generated by particle-particle interactions, which favour rigid-body rotation of the cluster as a whole. In addition, the particles at the boundary have velocity $u_R >  u_0$. This higher velocity at the boundary captures the experimental fact that although each bulk layer has two neighbouring layers, the outermost layer has only one neighbour and therefore
is slowed down less by its neighbours. In the continuum, this balance of friction and self--propulsion leads to an equation for the velocity $\bfv (\bfr)$ of layer at radius $r=|{\bf r}|$ in the steady state:
\begin{equation}
\label{eq:tonertu}
\tilde\nu \nabla^2 \bfv (\bfr)  - \xi \left(\bfv (\bfr) - u_0 \hat{\mathbf{e}}_{\varphi} \right) = 0,
\end{equation}
which can be obtained by linearising the steady-state Toner-Tu equations~\cite{toner1995}.  For our small system, it is natural to discretise into the five particle layers around the central particle within the rotating cluster. With $n = 1,\ldots,4$, $v_n = v(r_n)$ for the azimuthal velocity in plane polar coordinates, we obtain,

\begin{equation}
\tilde\nu \left( [\Delta_r^2 v]_n + r_n^{-1}[\Delta_r v ]_n - r_n^{-2} v_n \right)
- \xi \left( v_n - u_0 \right) = 0,
\label{eq:dd-main}
\end{equation}

where $[\Delta_r v ]_n$ is the discrete radial difference operator (see Supplementary Information for more details).

Equation~(\ref{eq:dd-main}) represents a system of 6 linear equations for the layer velocities (including the two boundary conditions $v_0 = 0$ and $v_5$ = $u_R$), which we solve  simultaneously. The results are shown as the solid lines in Fig.~\ref{figN61}\textbf{c}, which show quantitative agreement with the experimental profiles at intermediate field strengths. At higher fields, we obtain only qualitative agreement but observe, importantly, the inversion in flow profile. Our experimental discovery of flow field inversion is reproduced in a model with only two ingredients: a preferred active particle velocity, and friction between adjacent particle layers.

\section{Stick--slip behaviour}
The steady--state description that we have discussed so far assumes velocities that are constant in time. In reality, particle motion is intermittent, characterised by periods without relative motion between adjacent layers (``sticking'') interspersed with slipping events between adjacent layers. Slipping is associated with the creation and annihilation of localised defects in the hexagonal order. To identify these defects, we introduce the local bond--angle distortion parameter, $\Theta (t) = (1/\theta^{c}) \sum_{i}|\theta_{i} -\theta^{c}|$, computed at time $t$, for every simplex obtained from a Delaunay triangulation, where $\theta^{c} = \frac{\pi}{3}$. Here, $\Theta = 0$ represents equal bond angles in a perfectly ordered structure, whereas $\Theta = 1$ indicates a severe distortion to $\theta_i = \frac{2\pi}{3}$. In Fig.~\ref{figN61}\textbf{d}, we show the time--averaged radial profile of angular distortion $\Theta$ for a range of field strengths. As anticipated, $\Theta$ is largest under conditions that generate significant slipping between layers, both in the case that the outer layers move faster than the inner layers ($E \approx 1.4 \EQ$) and after the flow inversion when the inner layers move faster than the outer layers ($E \approx 2.3 \EQ$).

The non-monotonic behaviour observed in $\Theta$ is reflected in the layer--resolved bond orientational order parameter $\psi_6$, shown in Fig.~\ref{figN61}\textbf{e}. At low field strength, the system rotates rigidly, preserving hexagonal ordering and high $\psi_6$ in all layers. At the onset of slipping ($E \approx 1.4 \EQ$), hexagonal order is disrupted and $\psi_6$ is reduced. Hexagonality recovers at the higher field strengths ($1.8 \lesssim E \lesssim 2.0 \EQ$), as the system quickly relaxes to a locally hexagonal structure following each slip event (see Supplementary Movie 3). At the highest field strength ($E \approx 2.3 \EQ$), there is a significant drop in $\psi_6$ indicating fluidisation.

\begin{figure*}
\includegraphics[width=0.95\textwidth]{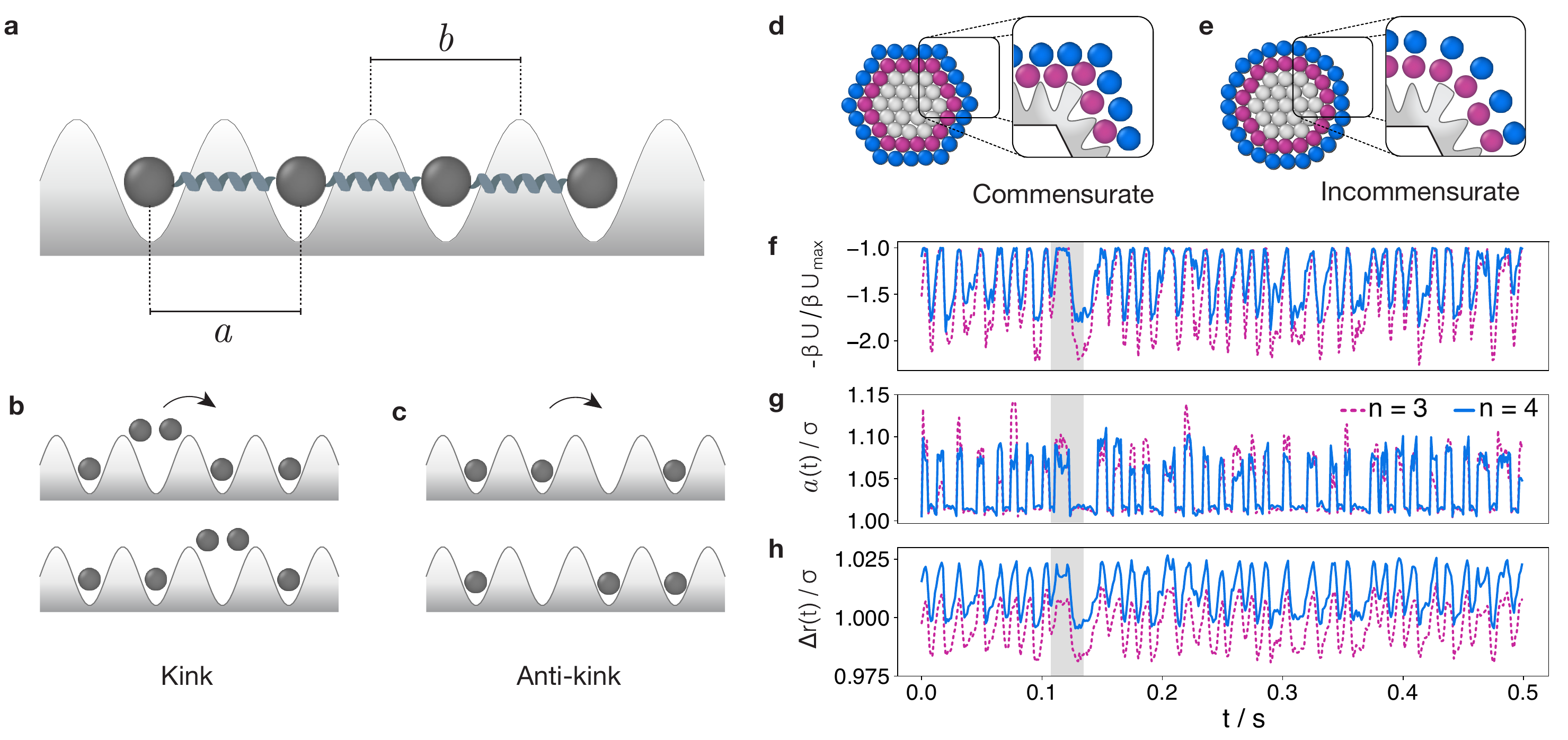}
\caption{\textbf{Frenkel--Kontorova behaviour in Quincke rotors.}
\textbf{a.}~Schematic representation of the Frenkel--Kontorova model. A slipping layer is modelled by a chain, comprising particles connected by springs of length $a$. The lattice periodicity is $b$, and the interaction between the static lattice and the sliding layer is given by a commensurate sinusoidal potential.
\textbf{b.}~Kink and \textbf{c.}~anti--kink representation. A kink forms when two particles are contained in the same well, producing a local compression of the chain. The opposite case of an anti--kink is a local expansion of the chain, which leads to a vacancy. For our confined geometry, the slipping layers are treated as chains forming
\textbf{d.}~commensurate and \textbf{e.}~incommensurate configurations.
\textbf{f.}~The evolution of the rescaled interaction potential $\beta U$,
\textbf{g.}~chain extension $a$, and \textbf{h.}~layer dilation follow the oscillations of $\psi_{6}$ in Fig.~\ref{figFK}\textbf{c}. In \textbf{d--f}, the blue lines are the outermost layer ($n=4$) and the dashed crimson lines are the adjacent internal layer, $n=3$. Shaded regions in panels \textbf{f--h} correspond to sequences shown in Fig.~\ref{figFK}\textbf{a--b}.
Here the field strength is $1.4\EQ$.
}
\label{figFK2}
\end{figure*}

We investigate slipping in more detail in Fig.~\ref{figFK}. To do so, we plot spatially resolved maps of $\Theta$ and $\psi_6$ as functions of time for a field strength of $1.4\EQ$ in Fig.~\ref{figFK}\textbf{a,b} (see also Supplementary Movie 3). During a slipping event, the two outer layers exhibit deviations from local hexagonal order at the vertices of the hexagon. This is reminiscent of slipping in driven assemblies of passive discs~\cite{williams2016}. We see that locally high values of the distortion parameter $\Theta$ are correlated with low values of $\psi_6$. These deviations from perfect hexagonal order are confined to the outer two layers. The solid--body nature of the interior protects the hexagonal order from any bond distortion. We illustrate the intermittent dynamics of $\psi_6$ showing sharp transitions between high and low values that are highly correlated between the two layers in Fig.~\ref{figFK}\textbf{c}.

\section{Slipping Mechanism: Frenkel--Kontorova Model in Periodic Geometry}
Slip in crystalline materials, in the form of layers sliding past each other, has long been understood in the context of the Frenkel--Kontorova model~\cite{chaikin}. This model refers to frictional dynamics in the transition between commensurate and incommensurate states when thermal fluctuations can be neglected~\cite{bohlein2012,ward2015}. Figure~\ref{figFK2}\textbf{a--c} illustrates the one--dimensional Frenkel--Kontorova model, in which a harmonic chain of particles with interactions modelled by springs of rest length $a$ is situated on a periodic potential corresponding to a substrate with period $b$. For a sliding chain in one dimension, the mechanism by which two lattices slip past each other involves the propagation of topological solitons known as kinks or anti--kinks. A kink corresponds to a localised compression of the chain that promotes a substrate spacing shared by two particles, as illustrated in Fig.~\ref{figFK2}\textbf{b}. Every kink propagation corresponds to the displacement of the local chain compression. On the other hand, anti--kinks are local extensions of the particle chain (Fig.~\ref{figFK2}\textbf{c}), which significantly reduce the friction between the particles.

Unlike the classical Frenkel--Kontorova model with an infinite chain, our confined system consists of periodic layers in the azimuthal direction, as shown in Fig.~\ref{figFK2}\textbf{d},\textbf{e}.  These illustrations show commensurate and incommensurate configurations during stick-slip dynamics and demonstrate that our system allows only for anti--kinks due to local extensions of the chain. The hard--core interactions between particles prohibit compression and the formation of Frenkel--Kontorova kinks \paddyspeaks{in these close-packed crystallites. However, in our system, anti-kinks correspond to stretching the system against the cohesive force between the particles, \emph{i.e.}, the electrohydroynamic interaction which binds the crystallite together.}

To interpret the slipping in our system in terms of the Frenkel--Kontorova model, we estimate the energetics of the process. Following Ref.~\cite{mauleon2020}, the interactions between the particles are modelled by an attractive Yukawa potential with a well depth $-\beta U(\sigma)=10$, where $\beta = (k_{\rm{B}}T)^{-1}$ is the inverse temperature. We use this to determine the potential energy of the assembly, which we plot as a function of time in Fig.~\ref{figFK2}\textbf{f}. The time--evolution is highly correlated with the structural order parameter $\psi_6$ shown in Fig.~\ref{figFK}\textbf{c}.

To measure the commensurability, we estimate the rest length $a$ as the nearest--neighbour separation in each slipping layer, with substrate period $b$ from the next layer internal to the slipping layer. In Fig.~\ref{figFK2}\textbf{g}, we see that these are anti--correlated with the potential energy. To consider the formation of anti--kinks, \emph{i.e.}, chain extensions, we determine the quantity $\sqrt{|\Delta \beta U| / (\kappa \sigma)}$ which is of order unity at the onset of slipping in the Frenkel--Kontorova model~\cite{chaikin}. To evaluate this quantity, we estimate the difference in potential energy between the slipping and non--slipping states $\Delta \beta U$ and the stiffness $\kappa$ for our system.

We find $\Delta \beta U$ by subtracting the potential energy per particle, $\beta U$, of the minima (corresponding to the commensurate states) from that of the maxima (corresponding to the incommensurate states, see Figs.~\ref{figFK2}\textbf{d--f}). We determine $\kappa$ by fitting a parabola to $\beta U(r)$ (see SFig. 1). We evaluate $\sqrt{|\Delta \beta U| / (\kappa \sigma)}$ for layers $n=2$ and $3$ ($0.107$) (c.f., measured value from Fig.~\ref{figFK2}\textbf{g} is 0.094) and between the outer layers $n = 3$ and $4$  ($0.075$) (c.f., measured value from Fig.~\ref{figFK2}\textbf{g} is 0.079). The geometry of our system enables slipping layers to dilate. We quantify the dilation through the deviation of each particle from their positions in the undeformed hexagonal case,  $\Delta r$.
This is shown in  Fig.~\ref{figFK2}\textbf{h} where we see that the dilation of the slipping layers is highly correlated with the potential energy and the chain extension (Figs.~\ref{figFK2}\textbf{f--g}).

While in the classic  Frenkel--Kontorova model, slipping is expected when $\sqrt{|\Delta \beta U| / (\kappa \sigma)} \sim 1$, here we find slipping at values around 0.10. This discrepancy can be resolved by observing (in Fig.~\ref{figFK}\textbf{c} and Figs.~\ref{figFK2}\textbf{f--h}) that the system undergoes slipping for only a small fraction of the time and, therefore, that $\sqrt{|\Delta \beta U| / (\kappa \sigma)} \ll 1$. If we evaluate the fraction of time the system spends slipping using the cutoff $\psi_{6} < 0.5$ (plotted as dashed lines in Fig.~\ref{figFK}\textbf{c}), we find that the system slips only $17\%$ of the time. Hence, the apparent discrepancy between the FK model and our calculations can be resolved due to the intermittent stick-slip dynamics which result from the interplay between dynamic frustration and crystalline order.

\section{Discussion and conclusion}
\vspace{-11pt}

We have investigated the behaviour of a model system of Quincke rollers under strong confinement. At low particle populations, this active system forms an apolar gas with a transition to a polar fluid for a higher number of rollers. Further increase of roller population leads to a hexagonally ordered packing.

For the magic number $\Nmax=61$, corresponding to a perfect hexagonal packing, these rollers exhibit a complex sequence of steady states as the activity is increased. At the level of individual layers, a phenomenological model based on self-propulsion and friction captures this behaviour. As activity is increased, the quiescent state gives way first to a state with rigid rotation, and then subsequently to a state with sliding layers. At lower fields, the outermost layer moves fastest but at higher fields, we experimentally observe a \emph{flow--inversion} transition to a state with the inner regions rotating faster than the boundary. Our model shows that this flow--field inversion is a direct consequence of the dynamic frustration between the linear velocity of self--propulsion and the collective angular velocity of a rotating rigid body.

At the particle--resolved level,
our experiments reveal
intermittent stick--slip dynamics between the layers.
We use measurements of the interactions between the particles~\cite{mauleon2020} to 
compare with predictions for continuous slipping from the Frenkel--Kontorova model. We find these parameters are much 
less than the criterion at which slipping is predicted (by a factor of ten). We interpret that this discrepancy is caused by our rotors slipping intermittently rather than continuously. Indeed we estimate that slipping occurs only 17\% of the time. At the highest activities, we observe a transition to continuous slipping which coincides with fluidisation.

A related passive system of a hard--sphere colloidal ``corral'' being driven from the boundary, like a tiny rheometer, has been investigated previously~\cite{williams2013,williams2016,williams2022}. Both the passive and active systems exhibit a fluidisation transition from a hexagonal configuration to a layered fluid. However, unlike the active system, the passive rotors show neither flow-field inversion nor intermittent slipping. This highlights how the new phenomenology characteristic of active solids results
in our current experiments from a combination of interparticle cohesion and activity.

Activity allows for solids in which complex dynamics emerge from simple ingredients. In our Quincke roller system, the formation of coherent active rotation due to confinement suggests a route towards the extraction of useful work from synthetic active matter.
We have demonstrated how cohesion and activity can be used to design soft nanomachines in which flow fields and intermittent dynamics can be manipulated at the microscopic scale.

\section*{Appendix A: Experimental Methods}

We employ a colloidal suspension of poly(methyl methacrylate) (PMMA) spheres, of diameter $\sigma = 3\mu\mathrm{m}$. Spheres are suspended in a low conductive solution of AOT surfactant 0.15 mM in hexadecane. Spheres are injected in a sample cell made of ITO-coated glass slides (Solems, ITOSOL12), separated by a layer of adhesive tape ($100\mu\mathrm{m}$ in thickness). Sedimentation occurs due to density mismatch with the solvent, and colloids form a quasi two--dimensional layer, as represented in Fig. 1 a in the main text. Quincke rotation is achieved with the application of a dc field $E$ using a potential amplifier (TREK 609E-6) connected to the sample cell.

Strong confinement is introduced by patterning the top electrode of the sampling cell. Circular regions of radius $R_{\mathrm{c}} \approx 5 \sigma$ are produced employing conventional photolithography methods. The thickness of the photoresist layer is $\approx 1 \sigma$, sufficient for charge screening on the top electrode. A non--zero electric current thus results only within the circular regions upon the application of the field $E$. This yields an electro--osmotic flow which traps rollers within the regions of interest. The origin of this flow is the transport of electric charges in the direction of the circular regions, which produces an inward flow at the bottom electrode~\cite{ristenpart2008}.

In the absence of the electric field, colloids are prone to escape the circular regions due to thermal motion. At low amplitudes of the field, e.g. $E < \EQ$, where $\EQ \approx 0.8\, \mathrm{V}\,\mu \mathrm{m}^{-1}$, a population of $N$ colloids is dragged towards the confining regions and self--assemble into clusters. Figure 1 b (main text) shows nine confining regions, each with a different population $N$. The maximum population observed corresponds to $\Nmax = 61$.

Quincke rollers are imaged using brightfield microscopy (Leica DMI 3000B) with a 10x magnification, and recorded at high speed (900 fps, Basler ACE) in regions of 128x128 px. The motion of individual rollers is reconstructed using a Python version of a common tracking algorithm~\cite{crocker1996}.

\begin{figure}[ht]
\centering
\includegraphics[width=0.35\textwidth]{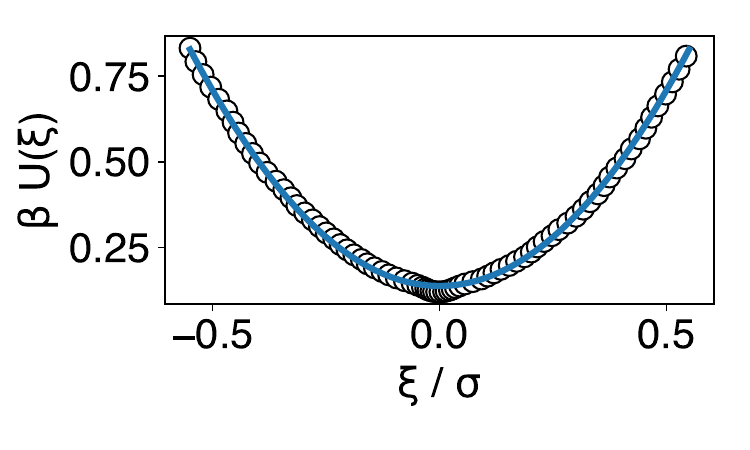}
\caption{Measured interaction potential $\beta U$ from experiments in the bulk \cite{mauleon2020} (symbols) and parabolic fit (solid line) from which a stiffness constant $\kappa$ is extracted.}
\end{figure}

\section*{Appendix B: Model of discrete hydrodynamics in circular confinement}

We combine an effective friction, which penalises relative motion between nearest neighbour particles, and self--propulsion with average speed $u_0$. For the circular domain, we assume that the motion is only in the azimuthal direction (i.e. the radial component of the velocity of each particle is zero) and that this velocity is on the same in each layer which we label by the number $n\in [1,4]$. Taking the particle diameter $\sigma$ (or the separation $a$), we can denote the radial position of the centre of each particle in layer $n$ by $r_n=\sigma(n-1)$. Given an azimuthal velocity $v_n = v(r_n)$, we can define discrete radial difference operator, $[\Delta_r v]_n = {v(r_n) - v(r_{n-1}) \over r_n - r_{n-1}}= \frac{1}{\sigma} \left(v_n-v_{n-1} \right)$. Applying it twice we get $[\Delta_r^2 v]_n = \frac{1}{\sigma^2} \left(v_{n+1}- 2 v_n + v_{n-1} \right)$. Balancing the local relative friction with self--propulsion leads to the equation,

\begin{equation}
\tilde\nu \left( [\Delta_r^2 v]_n + r_n^{-1}[\Delta_r v ]_n - r_n^{-2} v_n \right)
- \xi \left( v_n - u_0 \right) = 0,
\label{eq:dd}
\end{equation}

\noindent
where $\nu= \tilde \nu / \xi$ measures the local friction. The ``boundary conditions'' are $v_0=0$ and $v_5 = u_R$. Due to its position at the boundary with lower friction, we expect that the critical field is lower for the particles at the boundary than those in the bulk. Hence $u_R > u_0$. If the bulk critical field is $\EQ$ we take here that $u_0 =  A_0 \Delta E^\alpha$ where $\alpha=1/2$~\cite{bricard2013} for a single $\Delta E =(E-\EQ)$ and $u_R = B_R+ A_R \Delta E $.

This discrete dynamics can be mapped to a discretised version of the steady state of linearised Toner--Tu type equations for dry active matter for the local self--propulsion velocity $\bfv(\bfr)$ which in this setting is
\begin{equation}
\tilde\nu \nabla^2 \bfv (\bfr)  - \xi \left(\bfv (\bfr) - \bfu_0 \right) = 0,
\end{equation}

\noindent
where $\bfu_0= u_0 \hat{\bphi}$. For a circular domain, in plane polars we have $\bfr=(r,\phi)$ or $\bfr= r \rvec + \phi \phivec$ and $\bfv = v_r (\bfr) \rvec + v_\phi (\bfr) \hat{\bphi}$ with $0 \le r \le R$ and $\phi \in [0,2\pi]$. We consider the situation where $v_r=0$ and radially symmetric, such that $v_\phi(r)$ satisfies

\begin{equation}
\tilde\nu \left( \partial_r^2 v_\phi + r^{-1}\partial_r v_\phi - r^{-2} v_\phi \right)
- \xi \left( v_\phi - u_0 \right) = 0
\end{equation}

\noindent
where the boundary conditions are $v_\phi(0)=0, v_\phi(R)=u_R$.

The dynamics of equation (\ref{eq:dd}) can be expressed by the matrix equation

\begin{equation}
\sum_{m=0}^5 S_{nm} v_m = b_n \quad,
\end{equation}

\noindent
where $S_{mn}$ is a tridiagonal matrix and $b_n= -{\sigma^2 u_0 n^2 \over \nu}$.
%
%
Note that $R=5\sigma$. We fix the boundary conditions $v_0=0,v_5=u_R$ by setting the values of the 1st and 6th row of the matrix. The matrix is

\begin{widetext}
\begin{equation}
\bmS =\left(\begin{array}{cccccc}
-1 & 0 & 0 & 0 & 0 & 0 \\
0& -2-{\sigma^2 \over \nu} & 1 & 0 & 0 & 0  \\
0 & 2 & -7-4{\sigma^2 \over \nu} & 4 & 0 & 0  \\
0 & 0 & 6&  -16 - 9{\sigma^2 \over \nu} & 9 & 0  \\
0 & 0 & 0 & 12 & -29 - 16{\sigma^2 \over \nu} & 16  \\
0 & 0 & 0 & 0 & 0 & \lambda_5
\end{array}\right) \nonumber
\end{equation}

\noindent
where $\lambda_5= -25{\sigma^2 u_0 \over \nu u_R}$ is chosen to implement the boundary conditions. Then we can invert the matrix to solve for $v_n$.
Once we have $v_n$ we can obtain the angular velocity $\omega_n = v_n / r_n$. Parameters here are $a=1,\nu=10,A_0=1,A_R=0.7,B_R=0.5$, so that

\begin{equation}
{\boldsymbol \omega} =
\left( \begin{array}{c} 0 \\ (5760000 (0.5 + 0.7 \Delta E) + 5607396 \Delta E)/39975156 \\ (6048000 (0.5 + 0.7 \Delta E) + 3889008 \Delta E)/39975156 \\
(6499200 (0.5 + 0.7 \Delta E) + 2529372 \Delta E)/39975156 \\
(7137040 (0.5 + 0.7 \Delta E) + 1266484 \Delta E)/39975156
\end{array}\right)
\label{eq:WMatrix}
\end{equation}
\end{widetext}

\section*{Appendix C: Description of Supplementary Movies}

\textbf{Movie 1 ---} Experimental set-up. Movie shows multiple wells of diameter $R = 30 \mum$ containing different populations of Quincke rollers. Electric strength is $E = 1.4 \EQ$. Scale bar represents $100 \mum$. Movie is reproduced at 30 fps.
\\

\textbf{Movie 2 ---} Rigid behaviour at $E = \EQ$. Movie shows a roller population of $N = 61$ showing rigid body rotation. a) Experimental acquisition, and b) Shows guides to the eye for individual rollers located at different layers and over symmetry lines. Scale bar in a) represents $10 \mum$. Movie is reproduced at 6 fps.
\\

\textbf{Movie 3 ---} Slipping behaviour at $E = 1.4\EQ$. Same roller population showing steady-state slipping of the outermost layers. a) Experimental acquisition. b) Guides to the eye for individual rollers in different layers. c) Shows local hexagonal order $\psi_{6}$ and d) is the local distortion $\Theta$. Scale bar in a) represents $10 \mum$. Movie is reproduced at 6 fps.
\\

\textbf{Movie 4 ---} Onset of the flow inversion at $E = 2.0\EQ$. Movie shows the overtaking of internal layers past over the outermost layers. a) Experimental acquisition. b) Guides to the eye for individual rollers. c) Local hexagonal order $\psi_{6}$. Scale bar in a) represents $10 \mum$. Movie is reproduced at 6 fps.
\\

\textbf{Movie 5 ---} Fluid behaviour at $E = 2.3\EQ$. Movie shows the complete fluidisation of the structure, with every layer rotating with an independent angular velocity $\omega$. The flow inversion becomes more evident at this value of the field strength, where the internal layers rotate faster. a) Experimental acquisition. b) Guides to the eye for individual rollers at different layers. c) Local hexagonal order $\psi_{6}$. Scale bar in a) represents $10 \mum$. Movie is reproduced at 6 fps.
\\

\bibliography{rollingStones.bib}
\end{document}